# umx version 4.5: Extending Twin and Path-Based SEM in R with CLPM, MR-DoC, Definition Variables, Ωnyx Integration, and Censored distributions

Luis FS Castro-de-Araujo [1,2,*]; Nathan Gillespie[1]; Michael C Neale[1]; Timothy Bates[3]

[1] Virginia Institute for Psychiatric and Behavioral Genetics, Virginia Commonwealth University, p: +1 804 502-4074 P.O. Box 980126, Richmond, VA 23298-0126, USA. luis.araujo@vcuhealth.org .

[2] Dept of Psychiatry, The University of Melbourne, Austin Health, Victoria, Australia

[3.] Department of Psychology, University of Edinburgh, Edinburgh, UK

## Abstract

Structural Equation Modeling (SEM) is a flexible statistical technique with multiple applications, including behavioral genetics and social sciences. Building on the original design of the umx package, which improved accessibility to OpenMx by specifying a concise syntax, umx v4.5 extends functionality for longitudinal and causal twin designs while improving interoperability with graphical modelling tools such as Onyx. New capabilities include: classic and modern cross-lagged panel models; Mendelian Randomization Direction-of-Causation (MR-DoC) twin models incorporating polygenic scores as instruments; support for definition variables directly in umxRAM(); a workflow for importing paths from Ωnyx; a dedicated function for incorporating censored variables' data into models, particularly valuable in biomarker research; improved covariate placeholder handling for definition variables; sex-limitation modelling across five twin groups, accommodating quantitative and qualitative sex differences; and covariate residualization in wide- or long-format data. These new functionalities accelerate reproducible, reliable, publication-ready twin and family modelling, and integrated journal-quality reporting, thereby lowering barriers to genetic epidemiological analyzes.

# 1. Introduction

Structural equation modeling (SEM) is a statistical technique that allows hypothesis testing using complex models that include observed and latent variables. OpenMx (Neale et al., 2016) is a core engine for matrix- and path-based SEM, including multi-group twin models and definition variables, but scripting models from scratch may be a barrier for applied researchers due to its complexity. The R package umx was created to lower this barrier with a readable syntax, automatic labeling, start values, automatic plotting and reporting.

Since the 2019 paper (Bates, Maes, & Neale, 2019) which was based on umx version 1.8.0, package development has continued on CRAN and on GitHub (tbates/umx), with the inclusion new high-level models and usability improvements. Notable in v4.5 are: (i) umxCLPM for cross-lagged panel modeling; (ii) umxMRDoC for bidirectional causal inference with twin data and polygenic instruments; (iii) enhanced handling of definition variables directly within umxRAM(); (iv) workflows to bring Ωnyx-drawn (von Oertzen, Brandmaier, & Tsang, 2015) paths into umx; and (v) joint distribution analyses within the SEM specification. These additions reflect and leverage recent OpenMx, Ωnyx, and umx advances, and will be presented herein. Data preprocessing and reporting will be presented first, followed by high level functions.

# 2. Data preprocessing and reporting tools

## 2.1 Improvements for Twin modeling

A summary R method, which the user-facing umxSummary() function triggers, has been included since v1.8.0. All SEM functions, including the twin analysis functions, now produce a publication-ready report if the model object is passed to umxSummary(). It supports sorting parameters by type, reporting the fit indices, and provides interpretation guidelines for the fit indices' values, and can output to LaTeX or html for pasting into a word processor. Another new reporting tool is umxSummarizeTwinData(), which will produce a publication-ready table with means and group correlations (MZs, DZs, etc) for each variable stratified by twin pair type (all male, all female, and discordant pairs).

Modeling capabilities have also improved. Core functions such as umxACE now support adding covariates (both ordinal and continuous), multi-group specifications, and standardized reporting across twin ACE model, common pathway twin-model (CP), and simplex models. Additional features include support for gene–environment interaction (umxGxE_biv()), discordant twin designs (umxDiscTwin()), and power analysis tools (umxPower(), power.ACE.test()).

## 2.2 Integration and graphical specification.

Interoperability has improved, whereby umxRAM() now accepts lavaan syntax. This allows specifying a umxRAM model as a lavaan string. Exporting a model as a lavaan string, integrating umx into the space of lavaan users.

Ωnyx is a GUI tool for SEM in which models can be specified visually by drawing elements of a diagram, the resulting diagram can then be exported as OpenMx RAM and algebra code (von Oertzen et al., 2015). In umx v4.5, the Ωnyx exported path syntax can be directly read by umxRAM() or umxTwinMaker(). For twin models, umxTwinMaker() can transform Ωnyx-specified paths into an ACE specification, enabling a workflow from the graphical interface to advanced biometrical modeling of data from relatives.

The listing below is the exported Ωnyx OpenMx path code from the diagram in Figure 1. The naming of the A, C, and E variances must follow the pattern (a1, a2, a3, etc.) so that umxTwinMaker() can appropriately set the MZ and DZ covariance paths and constraints to specify a version of the model adapted for twin data. Note that the data-related lines from Ωnyx output need to be commented, as umxTwinMaker() will also handle data incorporation (in bold).

```
#
# This model specification was automatically generated by Onyx
#
# require("OpenMx");
# modelData <- read.table(DATAFILENAME, header = TRUE)
# manifests<-c("x1","x2","x3")
# latents<-c("icept","slope","a1","e1","a2","e2")
# model <- mxModel("umx2",
# type="RAM",
# manifestVars = manifests,
# latentVars = latents,
lgc_paths <-c(
mxPath(from="icept",to=c("x1","x2","x3"),                   free=c(FALSE,FALSE,FALSE),
value=c(1.0,1.0,1.0) , arrows=1, label=c("icept__x1","icept__x2","icept__x3") ),
mxPath(from="slope",to=c("x2","x3"), free=c(FALSE,FALSE), value=c(1.0,2.0) , arrows=1,
label=c("slope__x2","slope__x3") ),
mxPath(from="one",to=c("icept","slope"),    free=c(TRUE,TRUE),    value=c(1.0,1.0)    ,
arrows=1, label=c("const__icept","const__slope") ),
mxPath(from="a1",to=c("icept","slope"),    free=c(TRUE,TRUE),    value=c(1.0,1.0)    ,
arrows=1, label=c("a1__icept","a1__slope") ),
mxPath(from="e1",to=c("icept","slope"),    free=c(FALSE,TRUE),    value=c(1.0,1.0)    ,
arrows=1, label=c("e1__icept","e1__slope") ),
mxPath(from="a2",to=c("slope"),     free=c(TRUE),     value=c(1.0)     ,     arrows=1,
label=c("a2__slope") ),
mxPath(from="e2",to=c("slope"),     free=c(TRUE),     value=c(1.0)     ,     arrows=1,
label=c("e2__slope") ),
mxPath(from="x1",to=c("x1"), free=c(TRUE), value=c(1.0) , arrows=2, label=c("e") ),
mxPath(from="x2",to=c("x2"), free=c(TRUE), value=c(1.0) , arrows=2, label=c("e") ),
mxPath(from="x3",to=c("x3"), free=c(TRUE), value=c(1.0) , arrows=2, label=c("e") ),
mxPath(from="a1",to=c("a1"), free=c(FALSE), value=c(1.0) , arrows=2, label=c("VAR_a1")
),
```

```
mxPath(from="e1",to=c("e1"), free=c(FALSE), value=c(1.0) , arrows=2, label=c("VAR_e1")
),
mxPath(from="a2",to=c("a2"), free=c(FALSE), value=c(1.0) , arrows=2, label=c("VAR_a2")
),
mxPath(from="e2",to=c("e2"), free=c(FALSE), value=c(1.0) , arrows=2, label=c("VAR_e2")
),
mxPath(from="one",to=c("x1","x2","x3"), free=F, value=0, arrows=1)#,
#mxData(modelData, type = "raw")
);
  lgc_model <- umxTwinMaker(
    "lgc",
    paths = lgc_paths,
    mzData = mzData,
    dzData = dzData
  )
```

[Insert Figure 1 here]

The final OpenMx object lgc_model is a multiple groups model, with each submodel including the MZ and DZ specific data. This massively reduces scripting code for ACE models.

### 2.3 Joint Distribution Analyses: ICU Method

Users are often dealing with censored data - for instance bio-assay data often exhibit censoring, where values fall below a limit of detection (LOD). The Integrated Censored-Uncensored (ICU) method models – such below-threshold values as ordinal data and above threshold values as continuous data. The ordinal and continuous variables are treated as being perfectly correlated within persons. The expected covariance matrix is therefore singular, but, because any individual has only one of the two scores (ordinal or continuous), this need not cause problems with FIML estimation. The algorithm automatically trims the expected covariance matrix to contain only those rows and columns corresponding to the observed data vector. ICU modelling is a special case of censored data analysis for a pair of observed variables, where one variable is a binary with value falling below some threshold of detection and a second observed variable receives entries for above-threshold values. If there is an entry in one of these pairs of columns, the corresponding column has an NA. In umx, ICU modeling is supported via xmu_make_bin_cont_pair_data(), which prepares paired ordinal-continuous data for SEM analyses of data from either twins or unrelated individuals splitting them at a censoring point. The code listing below will result in two new variables with the bin and cont suffixes (mpgbin, mpgcont). Values of mpg below the chosen censoring point (0.001 in the example) are coded in mpgbin as an R ordered factor level labelled <low> (the variable can take two values: <low> and <high>), and values above 0.001 are coded NA in mpgbin. Conversely, the mpgcont variable is coded as NA if the mpg value is below 0.001, but as the actual mpg value if mpg>0.001. The matching diagram (RAM) specification for a twin example can be seen in Figure 2.

```
df = xmu_make_bin_cont_pair_data(uncensored_data,
                        vars = c("X1"),
                        censp = 0.001,  # the LOD
                        suffixes = c("_T1", "_T2")) # if twin data
```

[Insert Figure 2 here]

Once the data are loaded into the model, the user must fix the threshold for the binary variable at the LOD.

```
model$MZ$deviations_for_thresh$values[, c("X1bin_T1","X1bin_T2")]<- 0.001
model$MZ$deviations_for_thresh$free[, c("X1bin_T1","X1bin_T2")]<- FALSE
```

This approach reduces the need for data transformation, as it uses information available (threshold level) provided in the blood essay.

## 2.4 Covariate Placeholder for Definition Variables

Definition variables are additional variables that can be used directly in the specification of expected means or covariances. A common use case is ordinal data, where definition variables can be used to model the effects of covariates on outcomes of interest at the latent trait level. However, when using definition variables, rows with one or more missing definition variables are row-wise deleted.  As it is well-known in linear regression, row-wise deletion has adverse consequences for parameter estimates when data are either missing at random or not in the Little & Rubin (2014) sense.  This problem is exacerbated in the case of data collected from relatives, where, e.g., one twin has data on both the dependent variables of interest, and on their definition variables, but their cotwin has neither. The missing data on their cotwin's definition variable would force the exclusion of all the pair's data from the analysis, which is undesirable for two reasons.  First, there is waste of the data on the first relative. Second,  including all valid data may help to correct estimates for volunteer bias (Neale & Eaves, 1993).

The function xmu_update_covar() adds a placeholder to the covariate cell whenever the cotwin has information on that row for the same covariate, ensuring that as much information as possible is retained. The placeholder chosen is the numeric value 99999, which will be interpreted by the optimizer as an outlier and not affect the row's likelihood calculation.  In the event of a coding error (as might occur when processing such data manually), such as using 99999 for the covariate value when there are observed dependent variables for the cotwin, extreme parameter estimates and model-fitting issues are likely to be found, flagging the error to the user.

```
data(docData)
```

```
df = docData
# Add some missing data
df$varA1_T1[1:5] <- NA
df <- xmu_update_covar(df, covar = "varA1", pheno = "varB1")
head(df)
>  zygosity varA1_T1 varA2_T1 varA3_T1 varB1_T1 varB2_T1 varB3_T1
> 1 MZMM     99999    0.409    -0.449   NA        2.580    1.467
> 2 MZMM     99999    -0.765   -1.583   NA       -0.361    0.087
> 3 MZMM     99999     0.024    1.368   NA       -0.859   -0.967
> 4 MZMM     99999     1.048    1.069   NA       -0.433    0.307
> 5 MZMM     99999     1.393    2.719   NA        0.683    0.715
> 6 MZMM     -0.672    0.918    1.467  -1.388   -0.555    0.494
```

## 2.5 Residualizing Covariates in Long and Wide Data: umx_residualize

umx_residualize() provides a concise, reliable way to residualize one or more dependent variables (DVs) on covariates, returning the original data frame with those variables replaced by their residuals. It supports (i) a familiar rbase formula interface and (ii) fully supporting residualisation on wide-format twin data via suffixes automatically applying the same residualization to _T1, _T2, etc. columns. Internally, it wraps lm(..., na.action = na.exclude) and writes residuals back "in place", saving boilerplate model setup and assignment code.

Residualization is appropriate for continuous outcomes when you want to partial out exogenous covariates before SEM/twin modeling (e.g., age, sex, scanner). For ordinal outcomes, definition variables are to be preferred, as residualization is not defined on the latent liability scale used by threshold models. OpenMx handles ordinals via thresholds and modeling the effect of definition variables on the latent trait, rather than regressing them out prior to analysis. Notice, however, that pre-residualizing covariates or using definition variables comes with the assumption that those covariates are exogenous and measured without error; if that assumption fails, bias will propagate to downstream parameters.

The interface to umx_residualize() is designed to be consistent with the formulae used in lm() and other R packages. The primary operators are ~ for regressed on, + for main effects, and * for interactions, e.g., y ~ age * sex.

```
# (1) Residualize a single DV (mtcars example)
res1 <- umx_residualize("mpg", covs = c("cyl", "disp"), data = mtcars)

# (2) Formula interface, including non-linear terms and interactions
res2 <- umx_residualize(mpg ~ cyl + I(cyl^2) + disp, data = mtcars)

# (3) Residualize multiple DVs at once
res3 <- umx_residualize(var = c("mpg", "hp"), covs = c("cyl", "disp"), data = mtcars)

# (4) Residualize *wide* twin data by suffixes
tmp <- mtcars
```

```
tmp$mpg_T1  <- tmp$mpg_T2  <- tmp$mpg
tmp$cyl_T1  <- tmp$cyl_T2  <- tmp$cyl
tmp$disp_T1 <- tmp$disp_T2 <- tmp$disp
tmp <- umx_residualize(var = "mpg", covs = c("cyl", "disp"),
                       suffixes = c("_T1","_T2"), data = tmp)
```

This example uses the common "mtcars" dataset for accessibility to the widest set of users, but twin modellers should see that the same flexibility applies to wide twin data by simply setting the twin suffix parameter.

## 3. High-level functions

### 3.1 Random Intercept Cross-Lagged Panel Model (RI-CLPM)

The RI-CLPM, introduced by Hamaker et al. (2015), addresses a limitation of traditional CLPMs: the conflation of within-person dynamics with stable between-person differences. Standard CLPM (Luis FS. Castro-de-Araujo, de Araujo, Morais Xavier, & Kanaan, 2023; Heise, 1970) assumes that cross-lagged effects reflect causal processes, but these estimates are biased if individuals differ in trait-like levels of the variables. RI-CLPM introduces random intercepts for each variable, capturing stable between-person variance. This isolates within-person fluctuations across time, so cross-lagged paths estimate dynamic processes rather than trait-like (between-person) confounds.

There are only a few causal inferential models available. Among these, the CLPM family of models are distinctive in that they include temporality.  Auto-regressive paths adjusts the state of a variable in a previous moment in time, and any remaining variance is either measurement error, innovation, or a causal effect in the subsequent occasion (the cross-lagged path).

umx provides a high-level function that simplifies CLPM specifications with and without RI. It also provides options to incorporate elements from instrumental variable (IV) methodology into the CLPM. Singh et al. (2024) reported that the inclusion of instrumental variables in the context of CLPM provides extra degrees of freedom for within-wave causal inference, along with the cross-lagged paths. More recently, Hamaker et al. (2015) model was extended with instrumental variables (IV-RI-CLPM, L. F. Castro-de-Araujo et al., 2025), allowing within-wave causal investigation in the multi-level context, i.e., separating trait-like and state-like variation. The IV-RI-CLPM model (Luis Fs Castro-de-Araujo et al., 2025), as well as standard CLPM (Heise, 1970), and RI-CLPM (Hamaker et al., 2015) specifications are available in umx v4.5.

```
data(docData)              # example panel data
dt <- docData[2:9]         # select 4 waves for X and Y
m_clpm <- umxCLPM(
```

```
  waves   = 4,
  name    = "Hamaker2015",
  model   = "Hamaker2015",
  data    = dt,
  autoRun = TRUE
)
umxSummary(m_clpm)
```

### 3.2 Mendelian Randomization and MR-DoC

Mendelian Randomization (MR) uses genetic variants as instrumental variables to estimate causal effects between phenotypes in the presence of background confounding. umx includes a basic MR function umxMR, but also more sophisticated models capable of incorporating measured genetic data in twin designs. MR-DoC extends MR by combining instrumental variables and the direction-of-causation (DoC) model (Castro-de-Araujo et al., 2023; Minică, Dolan, Boomsma, de Geus, & Neale, 2018). Because the cross-twin cross-trait correlations provide extra degrees of freedom, MR-DoC allows the estimation of direct pleiotropic paths or full background confounding. The model comes in two flavours, MR-DoC (Minică et al., 2018) can estimate a direct pleiotropic path from the instrument to the outcome, but requires fixing unique environmental variance correlations between exposure and outcome for identification. The second flavour, MR-DoC2 (Castro-de-Araujo et al., 2023), can estimate causal paths in both directions and adjust for all background confounding, however not allowing for direct pleiotropic path estimation. MR-DoC2 is also identified with non-twin siblings, with the necessary constraint of coalescing additive genetic (A) and shared environmental (C) variances into a single familial resemblance (F) component.

The package now provides variance components specifications of a simple DoC model, as well as the MR-DoC and MR-DoC2 extensions. The implementation permits easy model parsimony tests, and supports ordinal variables via a latent threshold liability scale.

```
# mzData/dzData should contain twin pairs + phenotypes & PRS columns
m_mrdoc <- umxMRDoC(
  pheno  = c("BMI", "SBP"),         # exposure, outcome
  prss   = c("PRS_BMI"),            # single instrument, mrdoc
  mzData = mz_df,
  dzData = dz_df
)
umxSummary(m_mrdoc)
```

### 3.3 Definition Variables

A definition variable in OpenMx is a row-specific value that modifies the model for that observation. Unlike fixed parameters, definition variables allow per-subject moderation of means, variances, or paths. In umx v4.5, definition variables can be specified directly in umxRAM() simply using umxPath(defn = 'defvar'). This avoids manual algebra specification and supports continuous moderators without grouping. Further references to the definition variable should include the def_ prefix, umxPath(from=”def_defvar”, to = “X1”). This syntax corresponds to a specification where the definition variable affects the means of X1.

Definition variables are the only appropriate option when working with ordinal variables, as you can’t residualize an ordinal value. In psychology and psychiatry research, ordinal variables often are the majority of the data available. Therefore, the availability of this feature in the engine, OpenMx, is fundamental for research in these fields.

Currently, only OpenMx and MPLUS (Muthén, 2011) offer definition variables in SEM. Its use has been fundamental in gene by environment moderation, where the most used method leverages a definition variable to estimate the moderating effect (Purcell, 2002). Umx has provided tools for that specification since inception and includes plotting facilities (Bates et al., 2019). With the defn syntax, such moderations can be concisely included in any RAM/path specification. For the twin design specification, umx v4.5 provides a utility for placeholder inclusion (discussed in section 2.4) to avoid deleting rows for which only one of the twins has data in the definition variable cell.

### 3.4 Multigroup Sex-Limitation Twin Models: umxSexLim

umxSexLim() implements multivariate sex-limitation twin models in a correlated-factors framework across five groups (MZ male, DZ male, MZ female, DZ female, DZ opposite-sex). It enables (a) Quantitative differences (sex-specific magnitudes of A, C, E) and (b) Qualitative differences (distinct factors operating in one sex but not the other) while preserving identification for the DZ opposite-sex group via cross-sex correlations.

umxSexLim offers three nested specifications: nonscalar sex limitation, scalar sex limitation, and homogeneity. Nonscalar Sex Limitation allows for quantitative and qualitative sex differences on either A or C variances, with sex-specific inter-trait correlations Ra, Rc, Re, and free male–female correlations in the DZ opposite sex (DZOS) group. Scalar Sex Limitation allows only quantitative differences (distinct male/female paths), but a single set of Ra, Rc, Re shared across sexes. Homogeneity corresponds to the baseline ACE assumption (equal A, C, E structures across sexes; means are free to differ).

Due to limits in degrees of freedom in the classical twin study, qualitative differences can be modeled for only one of A or C at a time (A_or_C = "A" or "C"). The function defaults the A (dzAr) or C (dzCr) variance correlations to .5 and 1, respectively, which can be changed to test alternative models (e.g., dzCr = .25 for ADE). Assumptions typically include equating means/variances across birth order within zygosity groups. Finally, the umxSexLim() accepts ordinal phenotypes and configures the model object automatically.

```
library(umx)
# Example data: anthropometry (included with umx)
```

```
data("us_skinfold_data")

# Base names of traits (suffixes define twins, e.g., _T1/_T2)
selDVs <- c("tri", "bic", "caf")

# Split data by zygosity/sex (columns already wide with _T1/_T2)
mzm <- subset(us_skinfold_data, zygosity == "MZMM")
dzm <- subset(us_skinfold_data, zygosity == "DZMM")
mzf <- subset(us_skinfold_data, zygosity == "MZFF")
dzf <- subset(us_skinfold_data, zygosity == "DZFF")
dzo <- subset(us_skinfold_data, zygosity == "DZOS")

# Fit a Nonscalar model with qualitative differences on A
m_nsA <- umxSexLim(
  name   = "Skinfold_Nonscalar_A",
  selDVs = selDVs,
  sep    = "_T",           # base names + suffixes (_T1/_T2)
  mzmData = mzm, dzmData = dzm, mzfData = mzf, dzfData = dzf, dzoData = dzo,
  A_or_C = "A",            # qualitative differences modeled on A
  sexlim = "Nonscalar",
  autoRun = TRUE
)

# Journal-ready summary (CIs, optional genetic/environmental correlations)
umxSummarySexLim(m_nsA, showRg = TRUE, report = "html")
```

## 4. Summary

The development of umx v4.5 focused on facilitating data management, longitudinal analyses, and causal inference in twin research. It provides CLPM and MR-DoC in a consistent, simple and functional interface. The software makes the use of definition variables more convenient, and enhances interoperability with the graphical SEM software Ωnyx. These features support rapid prototyping, transparent reporting, and robust testing/simulation of developmental and causal hypotheses.

## Figure labels

### Figure 1

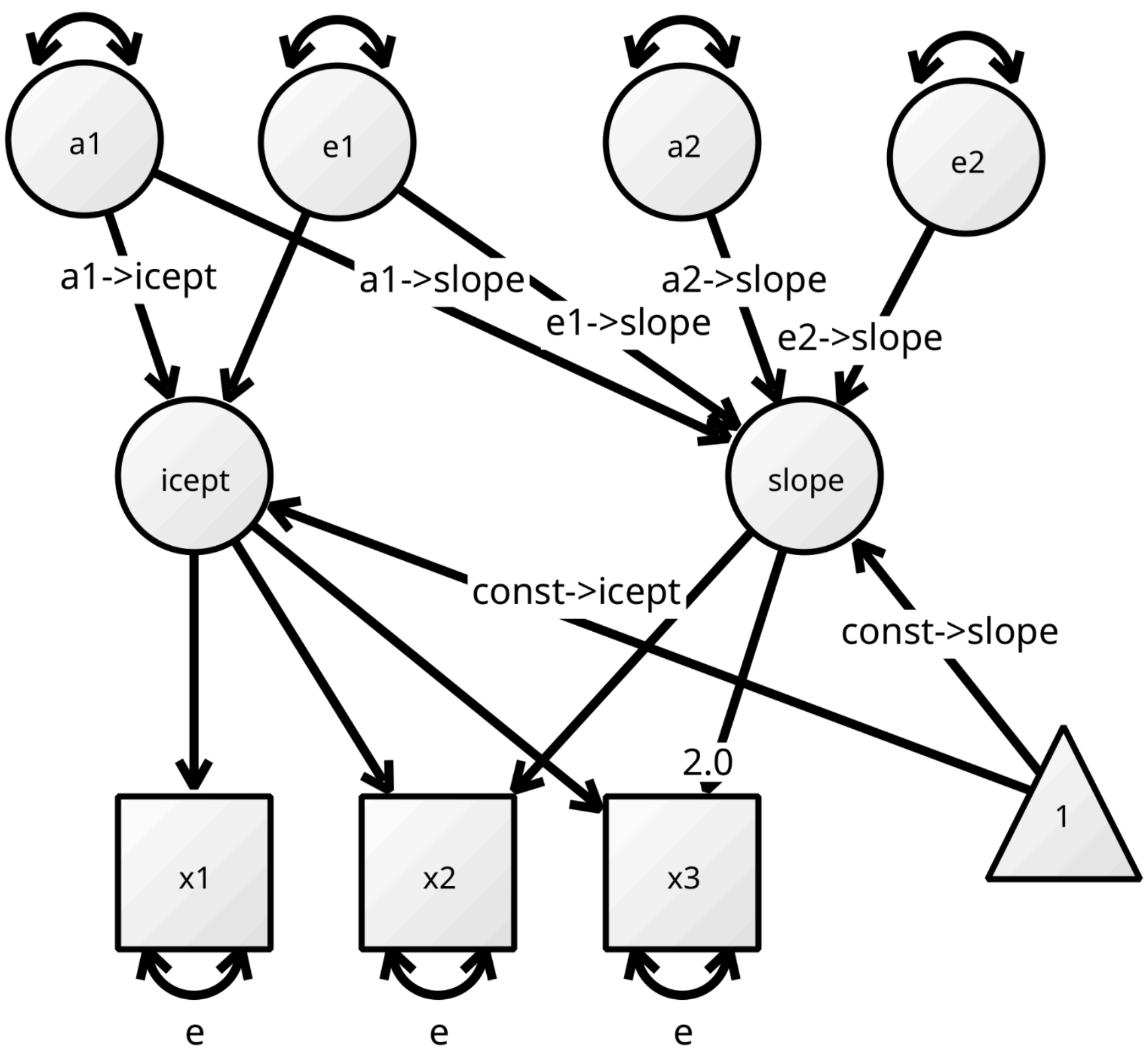

Ωnyx model diagram for a Cholesky AE specification. Notice the naming of the A and E variances should follow the pattern of a1, a2, a3, and so on, so that umxTwinMaker can set the remainder MZ/DZ paths and constraints for a twin model.

## Figure 2

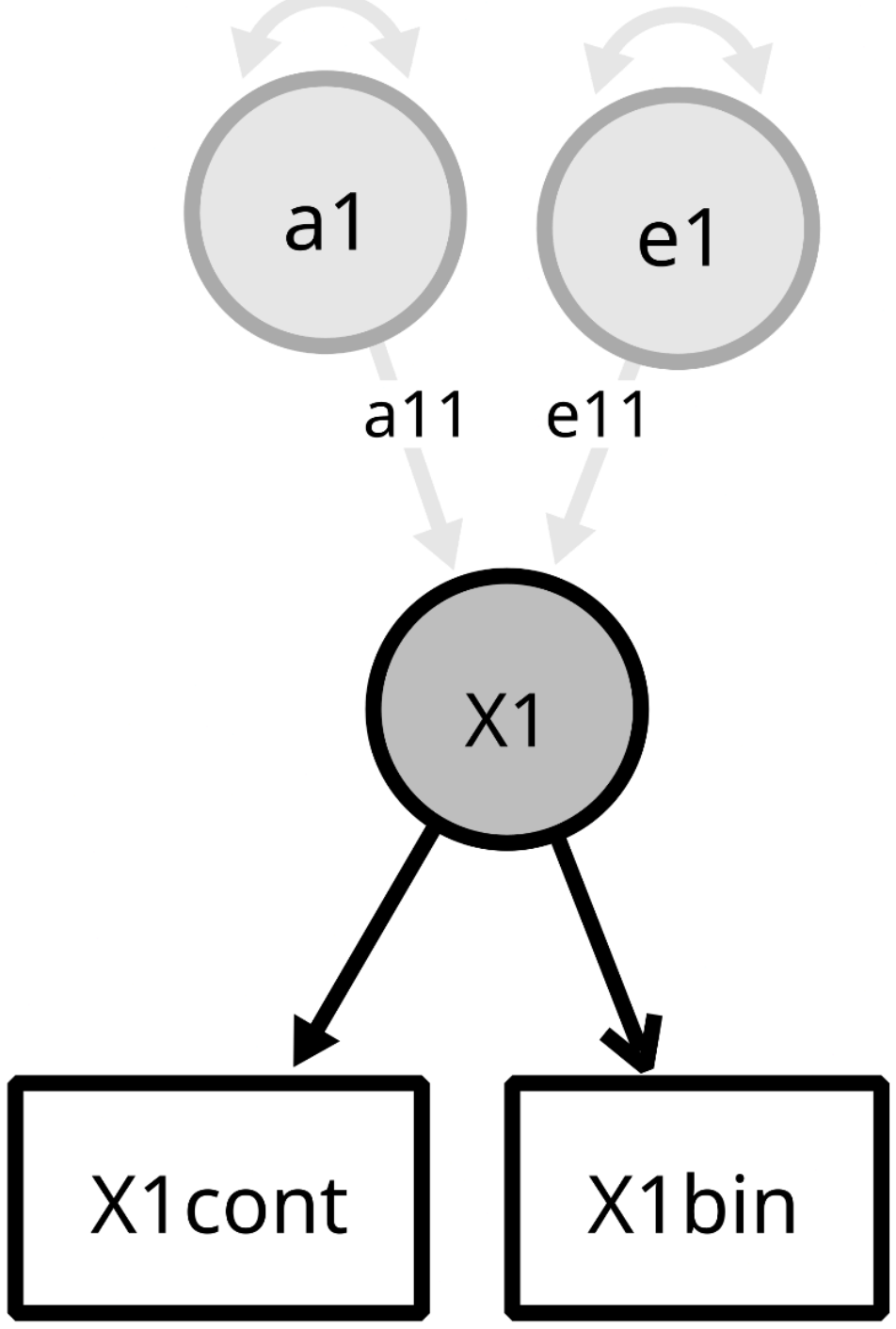

Ωnyx model diagram for a ICU specification. The X1 variable was split using xmu_make_bin_cont_pair() into X1cont and X1bin, the X1 latent variance now results from the joint-distribution analysis and is further split into A and E variances in a twin design.

## Acknowledgments

None

## Financial support

LFSCA is funded by NIH Grant No R01AG076838. MCN is funded by NIH grant DA-049867.

## Conflict of interest

Authors report no conflicts of interest.

## Ethical standards

Not applicable.